\documentclass[11pt]{amsart}
\usepackage{epsf}
\usepackage{amsmath}
\usepackage{amssymb}
\usepackage{cite}

\begin{document}
\def\vac{{\text{\O}}}

\parskip 2mm

\title{Stochastic cellular automaton for the coagulation-fission-process $2A \to 3A,\ 2A \to A$}
\author{Haye Hinrichsen}
\address{Theoretische Physik, Fachbereich 8,
     Universit\"at Wuppertal,
     42097 Wuppertal, Germany}
\date{July, 2002}

\begin{abstract}
We introduce an efficient cellular automaton for the coagulation-fission process with diffusion $2A \to 3A$, $2A \to A$ in arbitrary dimensions. As the well-known Domany-Kinzel model, it is defined on a tilted hypercubic lattice and evolves by parallel updates. The model exhibits a non-equilibrium phase transition from an active into an absorbing phase and its critical properties are expected to be of the same type as in the pair contact process with diffusion. High-precision simulations on a parallel computer suggest that various quantities of interest do not show the expected power-law scaling, calling for new approaches to understand this unusual type of critical behavior.
\end{abstract}
\maketitle

\section{Introduction}
%
Non-equilibrium phase transitions from fluctuating phases into one or several absorbing states continue to fascinate many researchers because of their universal properties~\cite{MarroDickman99,Hinrichsen00,OdorReview}. The ultimate goal is to categorize such phase transitions into a finite number of universality classes. So far several universality classes have been identified, the most important ones being directed percolation (DP)~\cite{DP}, the parity conserving (PC) class~\cite{PC}, voter-type transitions~\cite{Voter}, and the generalized epidemic process~\cite{GEP}. Searching for other universality classes, the pair contact process with diffusion (PCPD) $2A \to 3A$, $2A \to \emptyset$, also called annihilation-fission process, is one of the most promising candidates. In one spatial dimension the PCPD displays a novel type of critical behavior which does not seem to belong to any of the known universality classes. In contrast to DP and PC transitions, where individual particles can generate offspring, the PCPD is a {\em binary} spreading process, i.e., two particles have to meet in order to create a new one.

The possibility of a non-trivial critical behavior in binary spreading processes was already pointed out by Grassberger in 1982~\cite{Grassberger82}. Nethertheless it took another 15 years before the PCPD was investigated systematically for the first time by Howard and T\"auber~\cite{HowardTauber97}. Their work was motivated by the observation that the annihilation process $2A\to\emptyset$ produces anticorrelations among the particles (emerging as `imaginary' noise in the Langevin equation), while the fission process $2A\to3A$ generates positive correlations (`real' noise). Combining the two reactions they were able to show that the interpolation between `real' and `imaginary' noise leads to a non-equilibrium phase transition. However, the corresponding bosonic field theory turned out to be non-renormalizable. This is partly due to the fact that in the bosonic formulation the density of particles diverges in the active phase so that perturbative methods cannot be applied.

\begin{figure}
\epsfxsize=90mm
\centerline{\epsffile{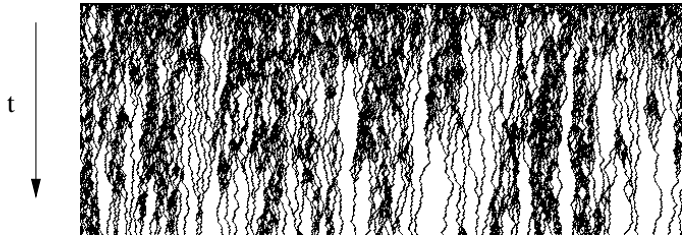}}
\caption{
\label{FIGDEMO}
Critical coagulation-fission process in 1+1 dimensions starting with a fully occupied lattice.
}
\end{figure}

Carlon~{\it et al.}~\cite{Carlon01} were the first to consider a fermionic lattice model of the PCPD, in which the local density of particles is bound by an exclusion principle. Their paper triggered a series of numerical studies~\cite{Hinrichsen01,Odor00,CyclicPaper,Odor01,PCPCPD,Odor02,Odor02b} and released a still ongoing debate concerning the critical behavior at the transition. Even today, two years later, it is not yet clear to what extent the PCPD represents an independent genuine universality class. On the one hand, a fermionic field theory for the PCPD does not yet exist. On the other hand, density matrix renormalization techniques and Monte Carlo simulations turned out to be unreliable because of strong deviations from  power law scaling. Moreover, in some studies the estimated critical exponents seem to depend on the choice of the diffusion rate, leading {\'O}dor to the conjecture that the transition may be characterized by two independent universality classes for weak and strong diffusion~\cite{Odor00}. Reported estimates for the critical exponents are scattered over a wide range, depending on the numerical effort and the value of the diffusion rate. Nevertheless there is a consensus that the PCPD does exhibit a novel type of non-equilibrium phase transition with a very characteristic critical behavior. The transition is universal in the sense that the same characteristic critical properties can be observed in various other binary spreading processes such as in multi-component models~\cite{Odor02b}, parity-conserving variants~\cite{PCPCPD}, and especially the coagulation-fission process~\cite{HenkelHinrichsen01,Odor01}
\begin{equation}
2A \to 3A \,, \quad 2A \to A \,,
\label{CoagFiss}
\end{equation}
on which we will focus in the present work (see Fig.~\ref{FIGDEMO}). By analyzing this process we will implicitely assume that the conclusions can also be applied to the PCPD and other binary spreading processes, at least on a qualitative level.

What is the reason for the unusually strong deviations from power-law scaling in the 1+1-dimensional PCPD? Initially it was argued that the deviations may be due to logarithmic corrections at the upper critical dimension. However, \'Odor showed that the upper critical dimension of the process is $d_c=2$~\cite{Odor02}. For this reason ordinary logarithmic corrections at the upper critical dimension can be ruled out unless the system has several critical dimensions. 

In order to explain the observed phenomena phenomenologically, it was suggested that the PCPD can be interpreted as a DP process coupled with an annihilation process in a feedback loop~\cite{CyclicPaper}. Numerical simulations confirmed that such a cyclically coupled model displays indeed a similar characteristic  critical behavior at the transition. Based on these ideas Noh and Park~\cite{NohPark02} proposed that the transition in the PCPD might be characterized by {\it continuously} varying exponents depending on the value of the diffusion rate. As an explanation they suggested that the PCPD may be viewed as a DP process subjected to a marginal perturbation caused by long-term memory effects of the dynamics. 

Even more recently Park and Kim presented numerical evidence that the PCPD is characterized by a well-defined set of critical exponents provided that the rates for diffusion and annihilation (or coagulation) are tuned in such a way that the process without branching becomes exactly solvable~\cite{ParkKimNew}. Very recent studies even suggest that all PCPD variants eventually cross over to the same asymptotic critical behavior with a unique set of critical exponents~\cite{Kwangho}, meaning that the PCPD would indeed represent an independent universality class.

So far all explanations anticipate that the critical PCPD displays asymptotic power laws. However, in the present study we find that there is no power-law scaling at all, at least within the numerically accessible range. To this end we introduce a very simple and efficient stochastic cellular automaton model for the coagulation-fission process~(\ref{CoagFiss}). Since the model evolves by synchronous updates, it can be implemented easily on a parallel computer. This allows us to perform high-precision simulations exceeding the number of time steps used in previous studies~\cite{GezaPrivate}. Our numerical results suggest that even after one million time steps the scaling regime is not yet reached. Inspired by this observation we discuss the possibility that the process may display an extremely slow crossover to directed percolation. 

\section{Definition of the cellular automaton}
%
%
\noindent
The cellular automaton for the 1+1-dimensional coagulation-fission process investigated in the present work is defined as follows. As the well-known Domany-Kinzel model~\cite{DKModel}, it uses a tilted square lattice, where the sites are enumerated horizontally by a position index $i$ and vertically by a temporal index $t$. Since the lattice is tilted the position indices $i$ at even (odd) times $t$ are also even (odd). The sites of the lattice can be either active ($s_i(t)=1$) or inactive ($s_i(t)=0$). Each $t$ labels a horizontal row, specifiying the state of the model at time $t$. 

\begin{figure}
\epsfxsize=110mm
\centerline{\epsffile{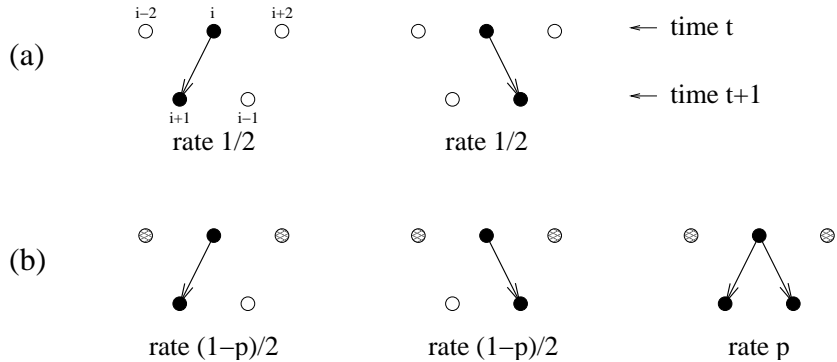}}
\caption{
\label{FIGRULES}
Dynamic rules of the cellular automaton in 1+1 dimensions. (a) Particles without a nearest neighbor perform a coagulating random walk by activating one of the adjacent sites in the new configuration with equal probablity. (b) If a particle has at least one nearest neighbor (indicated by grey color), both following sites are activated with probability $p$. Otherwise, the particle performs a coagulating random walk as in (a).
}
\end{figure}
The model evolves by parallel updates as follows. Assume that the state at time $t$ is given. In order to construct the new state at time $t+1$, we first clear all sites of the next horizontal row. Then a loop is executed which runs over all active sites $j$ of the previous configuration at time $t$. For each of them we first check whether its nearest neighbors $s_{j-2}(t)$ and $s_{j+2}(t)$ are occupied. Depending on their state the following local updates are carried out:
\begin{itemize}
\item[(a)] If both neighbors are inactive, the particle performs a coagulating random walk by activating either site $s_{j-1}(t+1)$ or $s_{j+1}(t+1)$ in the new configuration with equal probability. \\
\item[(b)] Otherwise, if at least one of the neighbors is active, another random number $z \in [0,1]$ is generated. If $z<p$ the two connected sites are activated by setting $s_{j-1}(t+1)=s_{j+1}(t+1)=1$, whereas for $z \geq p$ either site $s_{j-1}(t+1)$ or $s_{j+1}(t+1)$ is activated with equal probability in the same way as in (a).
\end{itemize}

\noindent
It is easy to see that these dynamic rules resemble the process $2A \to 3A$, $2A \to A$ combined with diffusion. Unlike in random-sequential models, it is not necessary to select particles randomly so that less random numbers have to be generated. A minimal pseudo code for the update algorithm is given in the appendix. In order to simulate the cellular automaton efficiently, further optimizations are necessary. Instead of using a static array describing a finite lattice with periodic boundary conditions, it is useful to store the positions of active sites at time $t$ individually in a dynamically generated list. Using such a list it is no longer necessary to perform a loop over all sites and to check whether they are active, instead the loop runs exclusively over the {\em active} sites, speeding up the simulation especially when the particle density is low. List-based algorithms are also advantageous in so-called dynamical simulations starting with a localized active seed~\cite{SeedSimulations}. Here the system size is practically infinite (limited only by the range of integer numbers $\pm 2^{31}$) so that finite-size effects can be excluded.

Moreover, the update procedure can be implemented easily on a parallel computer. To this end we note that the new state at time $t+1$ is initially set to zero and then constructed by an {\em unconditional} activation, i.e., we assign the value $1$ to certain sites irrespective of their state. For this reason the total update can be understood as a Boolean `OR' operation of all local updates within the loop. This means the local updates commute and thus can be processed in parallel blocks. Therefore, it is straight forward to implement the cellular automaton on a parallel computer with a ring geometry. Finally we note that the model can easily be generalized to higher dimensions in the same way as the Domany-Kinzel model, although the present work will be restricted to the 1+1-dimensional case.

\section{Numerical results}
%
%
\subsection{Homogeneous initial conditions}
%
%
\begin{figure}
\epsfxsize=160mm
\centerline{\epsffile{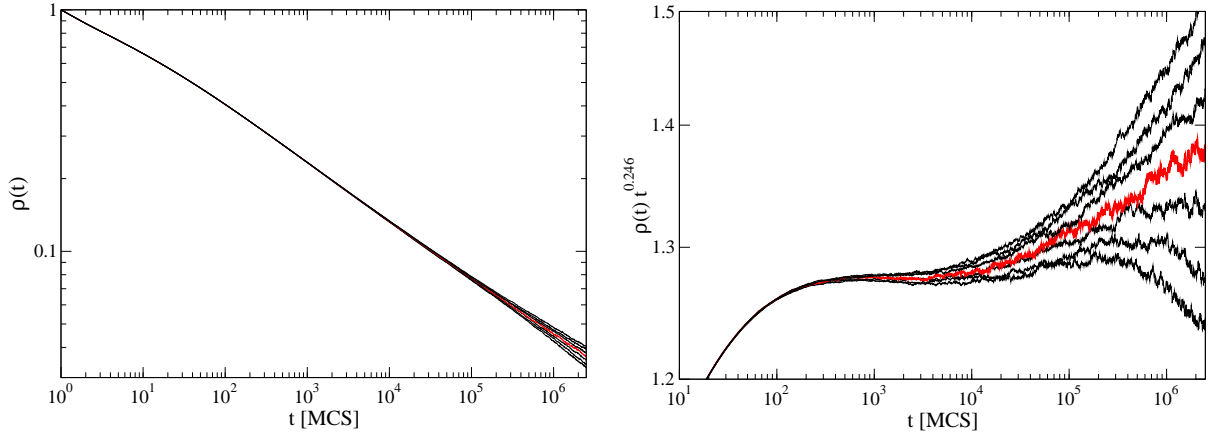}}
\caption{
\label{FIGDECAY}
Coagulation-fission process in 1+1 dimensions: Decay of the particle density close to criticality starting with homogeneous initial conditions. The left graph shows the raw data of seven independent simulations for $p$ running from 0.29992 (bottom) to 0.30004 (top) in steps of 0.00002. The right graphs shows the same data multiplied by $t^{0.246}$. 
}
\end{figure}
Fig.~\ref{FIGDECAY} shows the numerical results for the decay of the particle density close to the transition starting with a fully occupied lattice. In systems with power-law scaling the density is expected to decrease as $\rho(t) \sim t^{-\delta}$, where $\delta=\beta/\nu_\parallel$. The parallelized implementation allows us to simulate very large systems with $2^{21} \approx 2\times 10^6$ sites so that finite-size effects can be excluded. The simulation time of $2.5 \times 10^6$ updates exceeds the range of previous numerical works. Because of the unusually large system size it is sufficient to average the density of particles over $32$ independent runs. 

In order to estimate the critical threshold, we measured the decay for different values of~$p$. Plotting the raw data in a double logarithmic plot (left panel of Fig.~\ref{FIGEXPON}), there seems to be a reasonable straight line with a slope $-\delta \simeq -0.246(10)$ over almost four decades. However, if the measured values are multiplied by $t^{0.246}$ (right panel), it can be seen that the lines are actually curved. Note that the data sets for different $p$ are statistically independent so that the observed trend of the curvature cannot be attributed to correlated fluctuation effects.

The figure illustrates how difficult it is to determine the critical point. If the curves eventually veer down, we assume that the process is subcritical. Using this criterion we find that $p_c \geq 0.29998$. The exact critical point, which is presumably somewhere between $0.3000$ and $0.3002$, cannot be determined precisely since all data sets for $p \geq 0.29998$ show a non-vanishing positive curvature over at least three decades. Thus we have strong reasons to believe there {\em no} power-law scaling, at least in over the first six decades in time. In less extensive simulations one would have been tempted to interpret the broad plateau between $10^2$ and $10^4$ time steps as the onset of a power-law and to lower the percolation probability in order to get a straight line. The present numerical results show that this is not possible.

\subsection{Seed simulations}
%
%
\begin{figure}
\epsfxsize=160mm
\centerline{\epsffile{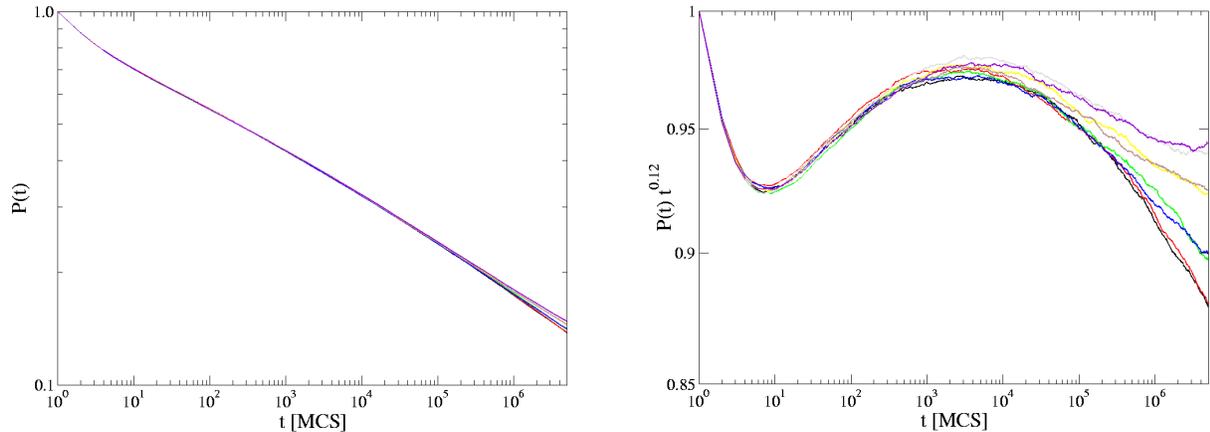}} 
\caption{
\label{FIGP}
Seed simulations. Left: Survival probablity as a function of time averaged over $6\times 10^6$ realizations for $p$ running from $0.29991$ (bottom) to $0.30012$ (top) in steps of $0.00003$. Right: The same data multiplied by $t^{0.12}$.
}
\end{figure}
\begin{figure}
\epsfxsize=160mm
\centerline{\epsffile{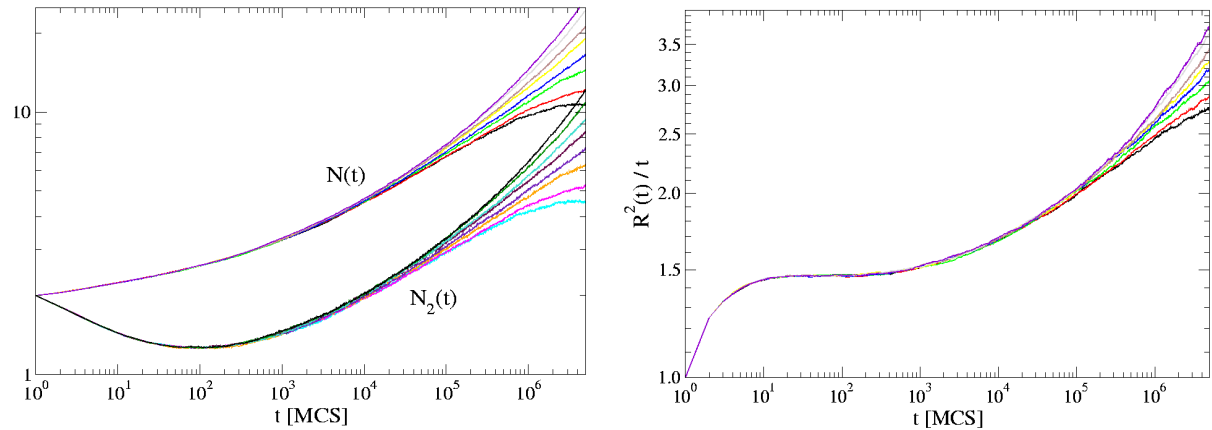}} 
\caption{
\label{FIGNR2}
Seed simulations. Left: Number of particles (upper curves) and number of particles with at least one active nearest neighbors (lower curves) as a function of time. Right: Mean square spreading from the origin averaged over surviving runs.
}
\end{figure}

Studying phase transitions into absorbing states it is customary to perform simulations starting with a localized seed~\cite{SeedSimulations}. The quantities of interest are the survival probability $P(t)$, the number of particles $N(t)$ averaged over all runs, and the mean square spreading from the origin $R^2(t)$ averaged over the surviving runs. In the standard scaling theory of phase transitions into absorbing states  these quantities are expected to obey asymptotic power laws of the form
\begin{equation}
P(t) \sim t^{-\delta'}\,,\qquad
N(t) \sim t^\eta \,,\qquad
R^2(t) \sim t^{2/z}\,.
\end{equation}
In the coagulation-fission process seed simulations are conceptually difficult. Since the absorbing state corresponds to a single diffusing particle, one has to start with a localized pair of particles. The survival probability is then most naturally defined as the probability that the system has not yet reached the absorbing state, i.e., the process is called `surviving' as long as there are at least two particles in the system. However, this definition is problematic in so far as it is no longer possible to divide the system into active sites and absorbing domains~\cite{GrassbergerPrivate1}. Irrespective of these difficulties, the survival probability $P(t)$ displays a similar behavior as the decay of the density in the case of homogeneous initial conditions. Plotting the raw data in a log-log plot one might expect a straight line with an approximate slope of $-\delta'=-0.12(1)$ (see left panel of~\ref{FIGP}). However, multiplying the data points by $t^{0.12}$ (right panel of Fig.~\ref{FIGP}) there is again a clear curvature over more than four decades. Thus, we can exclude power-law scaling, at least in the numerically accessible range.

Similar results are obtained for the other quantities. The left panel of Fig.~\ref{FIGNR2} shows the number of particles $N(t)$ averaged over all runs. In addition, the figure shows the average number of particles with at least one active nearest neighbor $N_2(t)$, i.e., the number of particles which can generate offspring. As previously observed in other studies of binary spreading processes in 1+1 dimensions, both quantities behave similarly and their quotient seems to tend to a constant. However, also in this case it is obvious that they do not follow a power-law. Likewise, the mean square spreading from the origin $R^2(t)$ does not display power-law scaling. As shown in the right panel of Fig.~\ref{FIGNR2}, after an initial transient there is horizontal plateau, where $R^2(t)$ increases linearly, indicating an intermediate diffusive behavior. After $10^3$ time steps, however, the curves veer up. This means that the effective exponent $z(t)$ decreases with time so that the process spreads {\em superdiffusively}.

\section{Does the PCPD belong to the DP class?}
\begin{figure}
\epsfxsize=150mm
\centerline{\epsffile{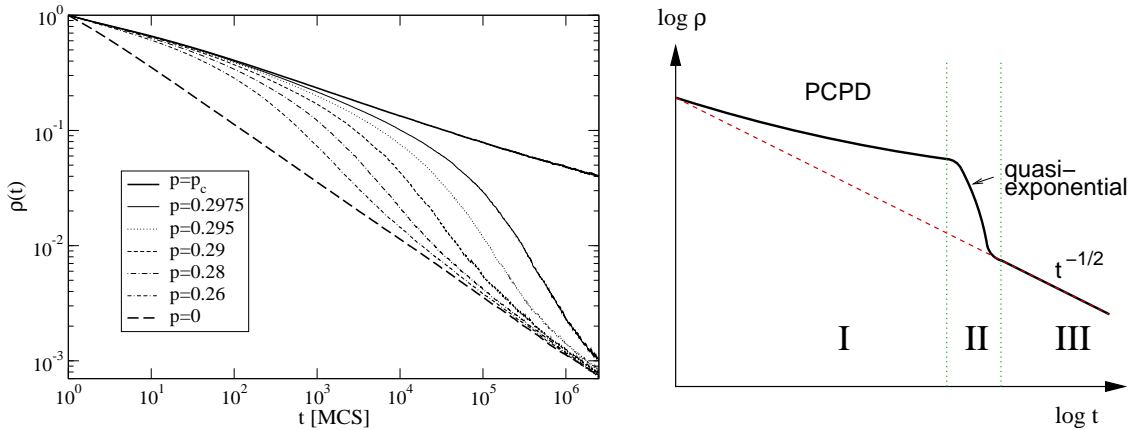}}
\caption{
\label{FIGEXPON}
Left: Decay of the density of particles in the inactive phase for various values of $p<p_c$. Right: Interpretation of the observed decay (see text).
}
\end{figure}

%
%
%
%
\begin{table}
\begin{tabular}{|c||c|c|c|}
\hline
exponent & $\quad t \approx 10^2 \quad$ & $\quad t \approx 10^6\quad $ & \;\; DP-value \;\; \\
\hline
$\delta$  & 0.25(1) & 0.22(1) & 0.159 \\
$\delta'$ & 0.11(1) & 0.13(1) & 0.159 \\
$\eta$    & 0.07(2) & 0.22(3) & 0.314 \\
$z$	  & 2.00(4) & 1.78(5) & 1.581 \\
\hline
\end{tabular}
\vglue 3mm
\caption{
\label{Table}
Effective exponents moving towards DP values as time proceeds.
}
\end{table}
\noindent
What is the origin for the unusual critical behavior observed in the coagulation-fission process. Why is it impossible to obtain clean power laws? One possibility, which has not been discussed so far, would be that the critical behavior of the PCPD and related models crosses over to DP after very long time. Several arguments support this speculative point of view:

\begin{enumerate}
\item
Unlike unary spreading processes, which generate offspring everywhere along the particle trajectories, the PCPD and related models are {\em binary} spreading processes, where solitary particles can move diffusively over long distances. As pointed out in Ref.~\cite{CyclicPaper}, the dynamics switches between two modes, namely, dense regions in space-time, where branching occurs frequently, and sparse regions, where solitary particles diffuse over long distances. The apparent contradiction that single particles diffuse (with a dynamic exponent $z=2$) while the whole process is superdiffusive ($z<2$) can only be resolved if the effective diffusion rate itself scales to zero under rescaling in the renormalization group sense. Therefore, the critical PCPD regarded on extremely large scales should look like a pair contact process without diffusion, which is known to belong to the DP universality class~\cite{PCP}.\\

\item
As shown in Table~\ref{Table}, all effective exponents move in the direction of the DP values as time proceeds. The same applies to the density exponent $\beta$ (not reported here). However, the accuracy of the numerical results does not permit a reliable extrapolation.\\

\item
As shown in Fig.~\ref{FIGNR2}, the effective dynamic exponent $z$ deviates only slightly from $2$. Between $10$ and $100$ time steps there is virtually no deviation. At $10^4$ time steps the deviation is less then
$8\%$ and after $10^6$ time steps it is still less than $12\%$. The small deviation makes it plausible why the diffusion constant changes only gradually under renormalization, leading to an extremely slow crossover.\\

\item
In the inactive phase a binary spreading process in 1+1 dimensions displays an algebraic decay $\rho(t) \sim t^{-1/2}$, whereas in DP the density is known to decay exponentially. This apparent contradiction can be resolved by observing that the decay of the particle density in a slightly subcritical PCPD can be divided into three different temporal regimes. In the first regime (region I in Fig.~\ref{FIGEXPON}) spreading and diffusion are still strong enough to sustain each other so that the particle density decays very slowly. After a characteristic time, however, the density decays quickly (region II) until it crosses over to an algebraic decay $t^{-1/2}$ (region III). Remarkably, the amplitude of this asymptotic power law does {\em not} depend on $p$. Consequently, the sudden jump in region II becomes more and more pronounced as the critical threshold is approached and may tend to a quasi-exponential decay in the limit $p \to p_c$.
\end{enumerate}
\noindent
In spite of these arguments the hypothesis of an eventual crossover to DP faces several problems. If it were valid in higher dimensions, it would be in conflict with \'Odors observation~\cite{Odor02} that the upper critical dimension of the PCPD is smaller than $4$. Moreover, it is not clear how the time reversal symmetry of DP is recovered in the asymptotic limit. In addition, the implicit assumption that all variants of binary spreading processes behave essentially in the same way has to be questioned. Finally, other high-precision simulations are reported that seem to confirm power-law behavior~\cite{Kwangho,GezaPrivate}.

\section{Summary}

\noindent
In the present work we have introduced an efficient cellular automaton model for the coag\-ulation-fission process with diffusion. Performing extensive simulations on a parallel computer, we find that the model does not exhibit power-law scaling in the numerically accessible range, instead the effective exponents (local slopes) show a continuous drift. As in various other studies, we may have been tempted to produce almost straight lines over the last few decades by adjusting the critical parameter, but the overall curvature of the data sets suggests that in this case the system would be slightly subcritical. For this reason we conjecture that the drift of the effective exponents will continue far beyond the numerically accessible range. We expect the same to be true for other binary spreading processes such as the PCPD. 

Observing that all effective exponents move towards the corresponding DP exponents (although they are still quite far away), one has to consider the possiblity that the system may cross over to DP after very long time. As a possible explanation we suggest that the diffusion rate may scale to zero in the renormalization group sense, driving the system eventually to a binary spreading process without diffusion, which is expected to display the critical behavior of DP. To this end it would be helpful to perform a real-space renormalization group study in order to see how the parameters, especially the diffusion rate, change under rescaling~\cite{Future}. \\

To summarize, there is a confusing variety of opinions concerning the critical behavior of binary spreading processes. Presently, the main viewpoints are:
\begin{itemize}
\item[-]
Binary spreading processes belong to a single
universality class with a well-defined set of critical exponents~\cite{ParkKimNew}.\\
\item[-]
Numerical results indicate that
there are two independent universality classes for low and high values of the diffusion rate~\cite{Carlon01,Odor00}.\\
\item[-]
The PCPD may be interpreted as a cyclically coupled DP and annihilation process~\cite{CyclicPaper}.\\
\item[-]
The PCPD can be regarded as a marginally perturbed DP process with continuously varying exponents~\cite{NohPark02}.\\
\item[-]
Binary spreading processes may cross over to DP after very long time (present work).\\
\item[-]
Finally, one cannot exclude the possibility that the PCPD and related models exhibit a weakly discontinuous transition~\cite{GrassbergerPrivate2}.
\end{itemize}
Further investigations are needed to find out which of these explanations is the correct one. Finally we note that the same questions arise in the context of a recently introduced triplet process with diffusion $3A \to 4A$, $3A\to \emptyset$~\cite{TripletProcess}. As in the case of the 1+1-dimensional PCPD, it may be possible that this process crosses over to DP as well after very long time.

\vspace{5mm}

\noindent {\bf Acknowledgements:} I would like to thank P. Grassberger, G. \'Odor, and K. Park for valuable discussions. The simulations were partly performed on the ALICE parallel computer at the IAI in Wuppertal. I would also like to thank B. Orth and G. Arnold for technical support.
\vspace{10mm}

\appendix\section{Minimal pseudo code}
%
\noindent
A minimal pseudo code (disregarding the boundary conditions) for the update procedure of the cellular automaton can be written as:

\begin{quote}
\small \tt
\begin{tabbing}
\noindent
integer L,T; \hspace{50mm} \=// system size and simulation time \\
integer s[L,T];		\>// the lattice\\
float   p;		 \>// the branching probability\\[2mm]
integer j;		\>// site index\\
integer  start = t mod 2; \>// select even/odd sublattice\\[2mm]

for j from 1 to L-1 do s[j,t+1]=0; 	\>// clear new state at time t+1 \\
for j from start to L-1 in steps of 2 do\>// loop over all active sites\\
	\hspace{1cm} 
	if s[j,t]==1 then\\
		\hspace{2cm}if random<0.5 then s[j-1,t+1]=1; 
			    else s[j+1,t+1]=1; \\
		\hspace{2cm}if s[j-2,t]+s[j+2,t]>0 and random<p then s[j-1,t+1]=s[j+1,t+1]=1;\\
	\hspace{1cm} end (if)\\
end (for)
\end{tabbing}
\normalsize \rm
\end{quote}

\vspace{4mm}

\newpage


\begin{thebibliography}{99}

\bibitem{MarroDickman99}
J.~Marro and R.~Dickman,
{\em Nonequilibrium phase transitions in lattice models},
Cambridge University Press, Cambridge (1999).

\bibitem{Hinrichsen00}
H.~Hinrichsen, Adv. Phys. {\bf 49}, 815 (2000).

\bibitem{OdorReview}
G. {\' O}dor, {\em Phase transition universality classes of
classical nonequilibrium systems}, eprint cond-mat/0205644.

\bibitem{DP}
W. Kinzel, {\em Percolation structures and processes}, Ann. Isr. Phys. Soc.
{\bf 5}, ed. G.~Deutscher and R.~Zallen and J.~Adler, Adam Hilger, Bristol (1983).

\bibitem{PC}
P. Grassberger, F. Krause, and T. von der Twer, Phys. A {\bf 17}, L105 (1984);
H.~Takayasu and A.~Y. Tretyakov,
Phys. Rev. Lett. {\bf 68}, 3060 (1992);
D.~ben Avraham, F.~Leyvraz, and S.~Redner,
{Phys. Rev.} E {\bf 50}, 1843 (1994);
J.~Cardy and U.~C. T{\"a}uber, Phys. Rev. Lett. {\bf 77}, 4780 (1996);
N.~Menyh\'ard,  J. Phys. A {\bf 27}, 6139 (1994);
M.~H. Kim and H.~Park,
Phys. Rev. Lett. {\bf 73}, 2579 (1994);
H.~Hinrichsen, Phys. Rev. E {\bf 55}, 219 (1997).

\bibitem{Voter}
T. M. Liggett, {\em Interacting particles systems}, Springer, Berlin (1985);
L. Peliti, J. Phys. A {\bf 19}, L365 (1986);
I. Dornic, H. Chat{\'e}, J. Chave, and H. Hinrichsen, 
Phys. Rev. Lett. {\bf 87} , 5701 (2001). 

\bibitem{GEP}
D. Mollison, J. R. Stat. Soc. B {\bf 39}, 283 (1977);
J. L. Cardy and P. Grassberger, J. Phys. A {\bf 18}, L267 (1985);
H. K. Janssen, Z. Phys. B {\bf 58}, 311 (1985);
P. Grassberger, H. Chat\'e and G. Rousseau, 
Phys. Rev. E {\bf 55}, 2488 (1997).

\bibitem{Grassberger82}
P. Grassberger, Z. Phys. {\bf B 47}, 365 (1982).

\bibitem{HowardTauber97}
M.~J. Howard and U.~C. T{\"a}uber, J. Phys. A {\bf 30}, 7721 (1997).

\bibitem{Carlon01} E. Carlon, M. Henkel, and U. Schollw{\"o}ck,
Phys. Rev. E {\bf 63}, 036101 (2001).

\bibitem{Hinrichsen01} 
H. Hinrichsen, Phys. Rev. E {\bf 63}, 036102 (2001).

\bibitem{Odor00} 
G. {\' O}dor, Phys. Rev. E {\bf 62}, R3027 (2000).

\bibitem{CyclicPaper}
H. Hinrichsen, Physica A {\bf 291}, 275 (2001). 

\bibitem{Odor01}
G. {\' O}dor, Phys. Rev. E {\bf 63}, 067104 (2001). 

\bibitem{PCPCPD}
K. Park, H. Hinrichsen, and I.-M. Kim, Phys. Rev. E {\bf 63}, 065103R (2001).

\bibitem{Odor02}
G. {\' O}dor, M. C. Marques, and M. A. Santos, Phys. Rev. E {\bf 65}, 056113 (2002). 

\bibitem{Odor02b}
G. {\' O}dor, Phys. Rev. E {\bf 66}, 026121 (2002). 

\bibitem{HenkelHinrichsen01}
M.~Henkel and H.~Hinrichsen, J. Phys. A {\bf 34}, 1561 (2001).

\bibitem{NohPark02}
J. D. Noh and H. Park, {\it Novel universality class of absorbing transitions with continuously varying exponents}, eprint cond-mat/0109516.

\bibitem{ParkKimNew}
K. Park and I.-M. Kim, {\em Well-defined set of exponents for a pair
contact process with diffusion}, to appear in Phys. Rev. E.

\bibitem{Kwangho}
K. Park, private communication (2002).

\bibitem{GezaPrivate}
In a very numerical recent numerical study, which goes even up to $10^8$ time steps, asymptotic power-laws have been identified. G. {\'Odor}, private communication (2002).

\bibitem{DKModel}
E. Domany and W. Kinzel, Phys. Rev. Lett. {\bf 53}, 311 (1984).

\bibitem{SeedSimulations}
P.~Grassberger and A.~de~la Torre, {Ann. Phys. (N.Y.)} {\bf 122}, 373 (1979).

\bibitem{GrassbergerPrivate1}
P. Grassberger, private communication (2001).

\bibitem{PCP}
I. Jensen, Int. J. Mod. Phys. B {\bf 8}, 3299 (1994).

\bibitem{GrassbergerPrivate2}
Speculations by P. Grassberger, private communication (2002).

\bibitem{Future}
K. Park and H. Hinrichsen, in preparation.

\bibitem{TripletProcess}
K. Park, H. Hinrichsen, and I.-M. Kim, Phys. Rev. E {\bf 66}, 025101R (2002).

\end{thebibliography}
\end{document}